# Is Planck's Constant $h$ a "Quantum" Constant?
# An Alternative Classical Interpretation

Timothy H. Boyer
Department of Physics, City College of the City University of New York

Abstract
Although Planck's constant $h$ is currently regarded as the elementary quantum of action appearing in quantum theory, it can also be interpreted as the multiplicative scale factor setting the scale of classical zero-point radiation appearing in classical electromagnetic theory. Relativistic classical electron theory with classical electromagnetic zero-point radiation gives many results in agreement with quantum theory. The areas of agreement between this classical theory and Nature seem worth further investigation.

## Introduction

Fundamental Constants

There are several fundamental constants of nature, including c, e, and $h$. According to the current physics literature, the constant c is the speed of light in vacuum, e is the charge on the electron, and $h$ is the elementary quantum of action. In this essay, we wish to raise the question as to whether $h$, Planck's constant, has been correctly identified. Specifically, we suggest that it may be useful to think of Planck's constant $h$ as the multiplicative factor setting the scale of classical zero-point radiation, rather than regarding $h$ as associated solely with quantum physics.

Planck's Constant

The story behind the introduction of Planck's constant is well-known and is often sketched out in the first chapters of textbooks of modern physics. Planck first introduced his constant $h$ in 1899 in connection with the spectrum of blackbody radiation. Blackbody radiation is the familiar radiation seen inside a glowing furnace. As the temperature of the furnace changes, the color spectrum of the radiation changes. Familiarly, we speak of objects as being red hot, or white hot, or blue hot. Physicists were trying to explain the spectrum which had recently come under careful experimental measurement. Planck identified two constants as needed to describe the spectrum of blackbody radiation. One constant was associated with Boltzmann's constant connecting energy and temperature. The second constant was unfamiliar but was related to Stefan's constant associated with the total thermal energy in the blackbody spectrum. Planck first made an inspired interpolation based upon his ideas of entropy. In 1900, Planck's interpolation fitted the experimental data extremely well, and, ever since, Planck's formula has been the accepted expression for the spectrum of thermal radiation. It was during his attempts to provide an underlying physical basis for his successful formula that Planck turned to the nonrelativistic classical statistical work of Boltzmann. Planck found that his formula appeared if he retained a finite elementary cell-size $h$ on phase space rather than taking the natural classical limit where the cell-size goes to zero. Thus within a year of its introduction, Planck's constant $h$ went from being merely an experimental parameter over to an association with an elementary unit of action.

Application of Planck's Constant

The blackbody radiation problem was hardly the only unsolved problem in physics at the end of the 19th century and the beginning of the 20th. Indeed Lord Kelvin[1] gave a lecture in 1900 entitled, "Nineteenth Century Clouds over the Dynamical Theory of Heat and Light." Kelvin's problems included the motion of the earth through the luminiferous ether and the decrease of specific heats at low temperature. Also, questions of atomic structure had become of interest. In the early 20th century, Planck's fundamental constant *h* with the units of action was found to be quite useful.

One of the first uses of Planck's constant was by Einstein in 1905. Einstein introduced the "revolutionary" idea of particles of light (later term "photons") with energy related to frequency ν, *E=hν*, to explain some aspects of Wein's law for blackbody radiation and also to account for the photoelectric effect. Particles of light were a complete departure from classical electromagnetic theory. In 1913, Bohr again proposed departing from classical electromagnetic theory by assuming that electrons in atoms had quantized energy levels $E_i$ where they did not radiate; and when electrons did radiate, the radiation involved sharp spectral lines with frequencies given by *ν=($E_i$-$E_j$)/h*. Thus stability came to the hydrogen atom by fiat, by changing the rules for electromagnetism at the atomic level. The structure of atoms was now associated with Planck's constant *h*. The radiation spectrum emitted by an atom was no longer associated with the classical analysis which works for radio waves and involves the time-Fourier transform of the electromagnetic field. Rather the atomic spectrum was reinterpreted as arriving as little balls of light marked with a discrete frequency ν and involving Planck's constant *h*. Bohr's simple rules fitted at least some aspects of the experimental data. The successes of the Bohr theory generated a conviction that a completely new "quantum" physics involving $h$ was required to understand nature at the atomic level. Heisenberg's quantum mechanics and Schroedinger's wave mechanics of the mid-1920s were the successful outcomes of this transformative conviction among physicists. Quantum theory was then applied to field theory, and the successes of quantum electrodynamics were spectacular. Indeed the quantum theory involving Planck's constant $h$ was triumphant. Every aspect of nature is now claimed to be traceable back to quantum phenomena involving Planck's constant *h*. Today we even introduce "virtual intermediate quantum states" to make sure that the observed facts can be fitted with quantum ideas. Quantum computers and quantum teleportation are among the latest quantum fads. In the course of a century, the quantum interpretation of Planck's constant *h* has gone from a surprising, vague hypothesis to unquestioned dogma.

## An Alternative Classical Understanding of Planck's Constant *h*

A Historical Irony

It is one of the ironies of the history of science that an alternative classical understanding of Planck's constant could have been introduced in 1900 but was not. Rather, the classical interpretation of Planck's constant has been developed slowly over the past half century by a tiny group of researchers.[2]

Classical Electron Theory in 1900
Today many physicists are unfamiliar with the classical electron theory of Lorentz which was in the forefront of research at the end of the 19th century. Classical electron theory involved point charges (electrons were identified in 1897) in nonrelativistic potentials which interacted through electromagnetic fields. Classical electron theory successfully described optical dispersion, birefringence, and even the normal Zeeman effect (for which Zeeman and Lorentz received the Nobel Prize in 1902). Classical electron theory involves three basic aspects: 1) Newton's second law with the Lorentz force for the motion of charged particles in electric and magnetic fields, 2) Maxwell's equations for the electromagnetic fields with electric charges as sources, and 3) boundary conditions on the equations. Lorentz's Columbia University lectures were published as *The Theory of Electrons*, and the second edition of 1915 is still available as a Dover paperback. In one of the Notes, Lorentz gives his explicit assumption regarding the homogeneous boundary condition on Maxwell's equations.[3] Lorentz assumed that the homogeneous solution of Maxwell's equation vanished; all radiation comes from the accelerations of charged particles at a finite time. This assumption still seems so natural, so obvious even to physicists today, that it never seems to be questioned in connection with classical physics.

The Casimir Effect
However, today there is strong experimental evidence that Lorentz' assumption is incorrect. One of the most transparent examples involves the Casimir effect,[4] the attraction between uncharged conducting parallel plates in the presence of random radiation. Any spectrum of random classical radiation will lead to a force between conducting parallel plates immersed in the radiation; each random radiation spectrum gives a different dependence for the variation of the force with the separation between the plates. In particular, the Rayleigh-Jeans spectrum of random thermal radiation gives a force between the plates which varies directly as the temperature T and inversely as the third power of the separation between the plates.[5] Now it is found experimentally at low temperatures that the force between the uncharged conducting plates does not vanish as the temperature goes to zero; rather, the force becomes independent of temperature. This suggests the possibility that there is random classical electromagnetic radiation present even at zero temperature. In other words, the universe contains classical electromagnetic zero-point radiation. Once we allow the possibility of classical electromagnetic zero-point radiation, we must ask what spectrum of random classical radiation would be suitable for zero-point radiation. It seems natural to say that the zero-point radiation should not pick out any particular inertial frame or length or time; it should be Lorentz invariant and scale invariant. It turns out that there is a unique spectrum or random classical radiation with these properties.[6] The spectrum has an energy $E$ per normal mode which is a constant times the frequency $v$ per normal mode, $E = const \times v$. We find that this random radiation spectrum leads to a theoretic force-dependence (as the inverse fourth power of the plate separation) which indeed matches that found experimentally at low temperatures. Fitting the classical theory to the experimental Casimir data, we find that the spectrum of random classical zero-point radiation has an energy $E$ per normal mode of frequency v given by $E = 3.31 \times 10^{-27} \times v$ where the constant is in erg-seconds. But this numerical constant is easily recognized as Planck's constant divided by 2. Thus within classical theory, Planck constant $h$ is not a quantum of action, but rather $h$ appears as the multiplicative constant setting the scale of random classical zero-point radiation.

Classical Zero-Point Radiation as Ambient Radiation
Classical zero-point radiation is similar to classical thermal radiation; it is the ambient radiation present in every part of the universe. Just as an experimenter who adjusts his equipment must make allowances for the thermal radiation which is present (and which he did not introduce), so also he must take account of the zero-point radiation. Just as the classical zero-point radiation leads to Casimir forces between conducting parallel plates, so too the classical zero-point radiation will interact with the experimenter's equipment, driving every part of it into zero-point motion.

Change in Classical Outlook Compared to 1900
The presence of classical electromagnetic zero-point radiation sharply alters the outlook on microscopic classical physics which existed in 1900. Indeed, Lorentz could have gone much further with his classical electron theory in the early years of the 20[th] century if he had changed his homogeneous boundary condition on Maxwell's equations to include classical electromagnetic zero-point radiation. However, there seems to have been very little serious consideration given to the possibility of classical zero-point radiation until the work of T. W. Marshall in the 1960s.[7] My own involvement with classical zero-point radiation began[8] in the late 1960s as an outgrowth of my work on the Casimir model of the electron and retarded dispersion forces.[9]

Charged Harmonic Oscillator in Zero-Point Radiation
Any classical charged harmonic oscillator will respond not only to coherent radiation and to thermal radiation but also to classical zero-point radiation. Indeed it is possible to solve analytically for the motion of a charged particle in a harmonic oscillator potential in classical electromagnetic zero-point radiation within the dipole approximation. It turns out that in the limit of small charge, the oscillator behavior is precisely that which is given in the quantum mechanics textbooks for the expectation values of products of any quantum operators (in symmetrized operator order). Indeed for free electromagnetic fields and linear oscillator systems, there is a general connection[10] between the classical theory with zero-point radiation and the quantum theory which appears in all the textbooks. It turns out that all the aspects of nature which can be described by quantized free fields or quantized harmonic oscillator systems can also be described equally well within classical theory including classical electromagnetic zero-point radiation. Thus Casimir forces, van der Waals forces, the specific heats of solids, diamagnetism, and the diamagnetism of a free particle can all be accounted for by classical theory which includes classical electromagnetic zero-point radiation.[6] It turns out that all of nonrelativistic classical statistical mechanics (which includes no zero-point energy) becomes valid merely as a nonrelativistic high-temperature limit.

Areas of Disagreement Between Classical and Quantum Theories
In the 1960s, it was hoped that the introduction of classical electromagnetic zero-point radiation into classical physics would provide an alternative hypothesis to quanta. The idea clearly worked for free fields and for harmonic oscillator systems. However, attempts to extend the idea to nonlinear nonrelativistic classical mechanical systems soon indicated that the hope was not fully justified. The calculation of rotator specific heats in the presence of classical zero-point radiation led to results which were at variance with those of quantum mechanics.[11] It seemed that the areas of agreement or disagreement between quantum theory and classical theory with zero-point radiation still needed to be determined.

Hydrogen Ground State in Classical Zero-Point Radiation
One place where it was hoped that classical zero-point radiation would be important is in the structure of the hydrogen atom. The ground state of hydrogen receives a natural qualitative explanation as an orbital Brownian motion if zero-point radiation is included within classical electrodynamics. The electron indeed radiates away its energy as it accelerates in its orbit around the proton. However, the electron also picks up energy from the random zero-point field. The balance between pick-up and loss should lead to an equilibrium situation for the electron in the hydrogen atom.[6] Although the linear stochastic differential equation for the charged harmonic oscillator in classical electromagnetic zero-point radiation is easy to solve analytically (in the dipole approximation), the motion of the electron in the Coulomb potential in zero-point radiation does not seem to allow analytic solution. However, Cole and Zou have done detailed simulation calculations[12] for the nonrelativistic motion of the electron and have shown that the probability distribution for the electron distance from the nucleus appears to closely resemble the probability distribution given by the Schroedinger ground state. The agreement is striking. Also, one should remember that Cole and Zou's classical simulation work allows no fitting parameters whatsoever.  The zero-point radiation is determined by agreement with the Casimir force between plates, and the electron charge and mass are known parameters.

Relativistic Mechanical Orbits in the Coulomb Potential
One important aspect of Cole and Zou's calculation is its comparison with the analytic work of Marshall and Claverie[13] who set up the hydrogen calculation in classical zero-point radiation using action-angle variables.  Marshall and Claverie never gave a probability distribution for the electron distance from the nucleus because it was concluded that the electron would pick up so much energy in the plunging orbits of small angular momentum that the atom would be ionized.[14]  Thus according to Marshall and Claverie, the problematic hydrogen atom of 1900 with its atomic collapse was being replaced by a hydrogen atom where the classical zero-point radiation was too strong.  However, most physicists are unaware that the plunging Coulomb orbits of small angular momentum are strongly altered by relativity. At large angular momentum, the relativistic mechanical orbits for a particle in a Coulomb or Kepler potential are very similar to the conic section orbits of nonrelativistic physics.  However, for small angular momentum (angular momentum less than $e^2/c$), the relativistic orbits are completely modified, and the orbiting particle plunges into the nucleus while *conserving energy and angular momentum!*[15] Thus Marshall and Claverie's orbits of ionization represent precisely the orbits where the nonrelativistic approximation is completely false.  We concluded that Cole and Zou have indeed given the correct ground state for hydrogen in the classical electromagnetic theory including classical zero-point radiation.

Importance of Relativity
Although the physicists of the early 20[th] century failed to realize the possibility of classical zero-point radiation, they and many other physicists have failed to appreciate the importance of relativity.  Special relativity is often regarded as a specialty theory needed only for high velocity particles.  It turns out that within classical physics, the blackbody radiation spectrum depends crucially on relativity.

Blackbody Radiation and Relativity
The blackbody radiation problem did not disappear with Planck's decisive interpolation and subsequent introduction of quanta in 1900.  Rather, the blackbody problem continued to occupy physicists from both classical and quantum perspectives all through the first quarter of the 20[th] century.  There were repeated derivations of the classical spectrum in concerted efforts to show that classical physics always produced the Rayleigh-Jeans result.  However, beginning in the late 1960s, it was shown repeatedly that classical physics including zero-point radiation led to the Planck result, not the Rayleigh-Jeans law.[8] Nevertheless, all these derivations left a nagging doubt about their validity because the scattering calculations using classical nonlinear nonrelativistic scattering systems pushed any spectrum of random classical  radiation toward the Rayleigh-Jeans law.[16]  Although the result had been conjectured for years, it was shown only recently that the classical zero-point radiation spectrum is indeed invariant under scattering by a *relativistic* scattering system.[17]  Contrary to what is suggested in so many physics textbooks, the Rayleigh-Jeans law for thermal radiation is associated not with classical physics but rather with *nonrelativistic* classical physics.  Nonrelativistic *classical* statistical mechanics cannot support the idea of zero-point energy (since kinetic energy is exchanged in classical mechanical collisions), and nonrelativistic mechanical waves are indeed associated with the Rayleigh-Jeans spectrum involving energy equipartition.  However, electromagnetism is a relativistic (not nonrelativistic) theory, and relativity is crucial for understanding blackbody radiation.  Only recently has a derivation of the Planck spectrum been given based upon relativistic classical physics involving conformal symmetry in non-inertial coordinate frames.[18]

# Classical and Quantum Physics in Noninertial Frames

## Transition to Noninertial Frames

Following the tremendous successes of quantum electrodynamics and the revival of interest in general relativity in the 1950s and 1960s, field theorists turned to questions involving quantum fields in noninertial coordinate frames. Fulling's Princeton doctoral thesis[19] discusses these questions and notes the non-uniqueness of the quantum vacuum state in non-inertial frames. For free fields in inertial coordinate frames, there is a close general connection between the expectation values of (the symmetrized products of) quantum fields and the average values of the corresponding classical fields.[10] However, this agreement disappears for noninertial frames. Irrespective of the spacetime metric, quantum field theory follows a prescriptive canonical quantization procedure to determine the field operators and vacuum state in any box, whether or not the box is at rest in an inertial frame or in an accelerating frame. Thus each box has its own quantum vacuum state. In complete contrast, classical zero-point radiation is based upon symmetry ideas. The natural extension of classical zero-point radiation to a noninertial frame involves the assumption that the correlation function for the random electromagnetic field involves only the geodesic separation between the spacetime points where the field is evaluated.[18] The contrast between the quantum and classical points of view has been work out for the situation involving constant proper acceleration through Minkowski spacetime.[20]

## The Unruh Effect

In the 1970s, it was discovered with some surprise that the time correlation function for a spacetime point accelerating through the quantum field theory vacuum in Minkowski spacetime seemed to involve the Planck frequency spectrum where the temperature was proportional to the acceleration. Indeed, this has been interpreted as indicating that an object accelerating through the Minkowski vacuum of quantum field theory experiences a thermal bath[21] at a temperature proportional to the acceleration (the Unruh effect); it is claimed that "accelerating stakes cook and eggs fry."

## Clash Between Quantum and Classical Predictions

However, since classical physics with classical zero-point radiation does not take its starting point for thermal behavior from Boltzmann's classical statistical mechanics, the classical interpretation is quite different from the quantum interpretation. According to the natural classical point of view, the accelerating object merely is in the ground state associated with the noninertial condition of the object. In the accelerating coordinate frame, the spatial correlation function for the radiation remains the familiar zero-point correlation function while the time correlation function is altered, just as accelerating clocks have their time altered.[18] Indeed the use of noninertial frames has provided the basis for a clear derivation of the Planck spectrum associated with the conformal symmetry of classical electromagnetic theory. According to some quantum theorists, if a box in the Minkowski vacuum state is gradually accelerated, it will go over to the Rindler vacuum state. However, if this same box is suddenly accelerated, the box will contain the Rindler excitations associated with a thermal bath. According to classical calculations, it does not matter how the box of zero-point radiation is accelerated; it always remains zero-point radiation.[20] Unfortunately, there is at present no experimental evidence regarding the thermal effects of acceleration so that at present we do not know which of these strikingly-different points of view is correct, the quantum or the classical.

# Rethinking Some Further Aspects of Physics

## Speculative Suggestions

The use of relativistic classical electron theory with classical electromagnetic zero-point radiation has already provided some interesting alternative points of view regarding some phenomena which are currently regarded as exclusively "quantum" phenomena. Both the classical and the quantum theories involve Planck's constant $h$, but with quite different interpretations. It might be fruitful to reexamine further aspects of modern physics in connection with the classical interpretation of $h$ as the scale factor for classical zero-point radiation. Among the phenomena which might be usefully examined are the following: the particle interference effects when particle pass through slits, the line spectra of isolated atoms, the behavior of matter at low temperatures, and the entanglement of microscopic systems.

## Particle Diffraction

It has already been shown that the presence of a classical zero-point radiation which pervades all space would produce Casimir forces between macroscopic objects and van der Waals forces between atoms and molecules. The presence of the atoms, molecules, dielectric materials, or conductors alters the correlation function for the classical zero-point radiation and so produces forces on the matter which interacts with the radiation. The presence of slits in a screen would also alter the correlation function for the classical zero-point radiation. Any particles which passed through the slits would interact with the altered zero-point radiation and so might move in a pattern which corresponded to wave-like behavior for the particles. Certainly, closing one of the slits would again alter the zero-point correlation function and so alter the pattern for particles passing through the slits.

## Sharp Line Spectra

The sharp line spectra of atoms have never been accounted for by classical physics. On the other hand, quantum theory has not provided much more detail than was suggested by Bohr's assumptions of 1913. At the same time, modern technology has made ever-smaller electrical antennas; and these antennas have been shown to operate along the same electromagnetic principles as the large-scale radio antennas. If classical electromagnetic zero-point radiation can indeed account for the excited states of atoms (something it has not done as yet), then presumably some sort of resonance phenomenon is involved. Indeed Cole and Zou[22] have shown that an electron in a Coulomb orbit has fascinating subharmonic resonances when the electron is driven by a circularly polarized plane wave. It may also be important to recall that an electron in a finite-size orbit picks up energy and loses energy at not just the fundamental frequency but at all the harmonics of the basic frequency. Just as it is hard to do relativistic calculations for the scattering of zero-point radiation by a relativistic scatterer, it may be hard to tease out the spectrum of the net emitted radiation for an atom when immersed in classical zero-point radiation.

## Matter at Low Temperature

Thermal radiation has velocity-dependent forces associated with motion relative to the coordinate frame in which the thermal radiation is isotropic. Zero-point radiation is isotropic in every inertial frame, and therefore does not give rise to velocity-dependent forces. At low temperatures, zero-point radiation (with its absence of velocity-dependent forces) gives a very different type of behavior for a collection of atoms and molecules than the behavior at high temperatures (where velocity-dependent forces are indeed present). Thus it may be useful to think in classical terms for situations currently described as involving superfluidity and Bose-Einstein condensation.

Entanglement and Classical Zero-Point Radiation
One might speculate that the idea of entanglement for quantum systems may have a connection to the fluctuations of zero-point radiation which will influence atomic systems.  Certainly the correlations between separated electric dipole harmonic oscillator systems can be described either in terms of quantum theory or in terms of classical theory with classical zero-point radiation.  Perhaps entanglement can be described in terms of the entanglement of the system with classical electromagnetic zero-point radiation.

## Closing Summary

Although quantum theory is the dominant theory of physics of our time, it may be useful to explore other theories in our effort to advance physics.  The interpretation of Planck's constant $h$ as a quantum of action proved enormously fruitful in the hands of physicists during the 20$^{th}$ century.  However, Planck's constant $h$ also has an interpretation as the multiplicative scale factor in classical electromagnetic zero-point radiation.  Relativistic classical electron theory which includes classical electromagnetic zero-point radiation has been shown to provide classical explanations for a number of phenomena which were previous thought to require quantum descriptions.  It is still an open question as to how much of Nature can be described in terms of classical physics which includes classical electromagnetic zero-point radiation.

Acknowledgement:  This essay was an entry in the FQXi essay contest of 2012.  I wish to thank Mr. Jonathan Ben-Benjamin, a student in my Electromagnetic Theory class, for bringing the contest to my attention.

References
1. Lord Kelvin, "Nineteenth Century Clouds over the Dynamical Theory of Heat and Light," Philosophical Magazine (ser. 6) **2**, 1–40 (1901).
2. A review of the work on classical electromagnetic zero-point radiation up to 1996 is provided by L. de la Pena and A. M. Cetto, *The Quantum Dice - An Introduction to Stochastic Electrodynamics* (Kluwer Academic, Dordrecht 1996).
3. H. A. Lorentz, *The Theory of Electrons* (Dover, New York, 1952). This is a republication of the 2$^{nd}$ edition of 1915. Note 6, p. 240, gives Lorentz's explicit assumption on the boundary conditions.
4. See, for example, Kimball A. Milton, *The Casimir Effect* (World Scientific, Singapore 2001).
5. T. H. Boyer, "Temperature dependence of Van der Waals forces in classical electrodynamics with classical electromagnetic zero-point radiation," Phys. Rev. A **11**, 1650-1663 (1975).
6. T. H. Boyer, "Random electrodynamics: The theory of classical electrodynamics with classical electromagnetic zero-point radiation," Phys. Rev. D **11**, 790-808 (1975).
7. T. W. Marshall, "Random electrodynamics," Proc. R. Soc A**276**, 475-491 (1963).
8. T. H. Boyer, "Derivation of the Blackbody Radiation Spectrum without Quantum Assumptions," Phys. Rev. **182**, 1374-1383 (1969).
9. T. H. Boyer, "Some Aspects of Quantum Electromagnetic Zero-Point Energy and Retarded Dispersion Forces," Harvard doctoral thesis 1968.
10. T. H. Boyer, "General connection between random electrodynamics and quantum electrodynamics for free electromagnetic fields and for dipole oscillator systems," Phys. Rev. D **11**, 809-830 (1975).
11. T. H. Boyer, "Specific heat of a classical, plane, rigid, dipole rotator in electromagnetic zero-point radiation," Phys. Rev. D **1**, 2257-2268 (1970).
12. D. C. Cole and Y. Zou, "Quantum Mechanical Ground State of Hydrogen Obtained from Classical Electrodynamics," Phys. Lett. A **317**, 14-20 (2003).
13. T. W. Marshall and P. Claverie, "Stochastic electrodynamics of nonlinear systems. I: Particle in a central field of force," J. Math. Phys. **21**, 1819-1925 (1980).
14. P. Claverie and F. Soto, "Nonrecurrence of the stochastic process for the hydrogen atom problem in stochastic electrodynamics," J. Math. Phys. **23** 753-759 (1982).
15. T. H. Boyer, "Unfamiliar trajectories for a relativistic particle in a Kepler or Coulomb potential," Am. J. Phys. **75**, 992-997 (2004).
16. T. H. Boyer, "Equilibrium of random classical electromagnetic radiation in the presence of a nonrelativistic nonlinear electric dipole oscillator," Phys. Rev. D **13**, 2832-2845 (1976).
17. T. H. Boyer, "Blackbody radiation and the scaling symmetry of relativistic classical electron theory with classical electromagnetic zero-point radiation," Found. Phys. **40**, 1102-1116 (2010).
18. T. H. Boyer, "The blackbody radiation spectrum follows from zero-point radiation and the structure of relativistic spacetime in classical physics," Found. Phys. **42**, 595-614 (2012).
19. S. A. Fulling, "Nonuniqueness of canonical field quantization in Riemannian space-time," Phys. Rev. D **7**, 2850-2862 (1973).
20. T. H. Boyer, "Contrasting Classical and Quantum Vacuum States in Non-Inertial Frames," arxiv 1204.6036 gr-qc (2012).
21. See the review by L. C. B. Crispino, A. Higuchi, G. E. A. Matsas, "The Unruh effect and its applications," Rev. Mod. Phys. **80**, 787-839 (2008).
22. D. C. Cole and Y. Zou, "Analysis of Orbital Decay Time for the Classical Hydrogen Atom Interacting with Circularly Polarized Electromagnetic Radiation," Physical Review E **69**, 016601(12) (2004); "Subharmonic resonance behavior for the classical hydrogen atomic system," Journal of Scientific Computing **39**, 1-27 (2009).